\documentclass[11pt]{article}
\usepackage{amsmath}
\usepackage{times}
\usepackage{graphicx}
\usepackage{color}
\usepackage{multirow}
\usepackage{libertine}
\usepackage[libertine]{newtxmath}
\usepackage[authoryear]{natbib}
\usepackage{float}
\usepackage{setspace}
\usepackage{cite}
\usepackage{rotating}
\usepackage{bbm}
\usepackage{latexsym}
\usepackage[hidelinks]{hyperref}

\newcommand{\captionfonts}{\footnotesize}
\makeatletter  
\long\def\@makecaption#1#2{%
  \vskip\abovecaptionskip
  \sbox\@tempboxa{{\captionfonts #1: #2}}%
  \ifdim \wd\@tempboxa >\hsize
    {\captionfonts #1: #2\par}
  \else
    \hbox to\hsize{\hfil\box\@tempboxa\hfil}%
  \fi
  \vskip\belowcaptionskip}
\makeatother   
 	
\begin{document}
\hspace{13.9cm}1

\ \vspace{-5mm}\\
{\LARGE Autoregressive Point-Processes as Latent State-Space Models: a Moment-Closure Approach to Fluctuations and Autocorrelations}
\ \\
{\bf \large Michael Rule$^{\displaystyle 1}$, Guido Sanguinetti$^{\displaystyle 1}$}\\
{\vspace{-0.5cm}$^{\displaystyle 1}$Institute for Adaptive and Neural Computation,
\\\vspace{-0.5cm}School of Informatics,
\\\vspace{-0.5cm}University of Edinburgh
\\\vspace{-0.5cm}10 Crichton St
\\Edinburgh EH8 9AB}
\\
{\bf Keywords:} Point processes, Moment closure, Mean-field
\\
%{\bf Running Title:} Moment-Closure of Autoregressive Point-Processes

\thispagestyle{empty}
\markboth{}{NC instructions}
\ \vspace{-10mm}\\
\singlespacing

%%%%%%%%%%%%%%%%%%%%%%%%%%%%%%%%%%%%%%%%%%%%%%%%%%%%%%%%%%%%%%%%%%%%%%%%%%%%
%Abstract
\begin{center} {\bf Abstract} \end{center}
Modeling and interpreting spike train data is a task of central importance in computational neuroscience, with significant translational implications. Two popular classes of data-driven models for this task are autoregressive Point Process Generalized Linear models (PPGLM) and latent State-Space models (SSM) with point-process observations. In this letter, we derive a mathematical connection between these two classes of models. By introducing an auxiliary history process, we represent exactly a PPGLM in terms of a latent, infinite dimensional dynamical system, which can then be mapped onto an SSM by basis function projections and moment closure. This representation provides a new perspective on widely used methods for modeling spike data, and also suggests novel algorithmic approaches to fitting such models. We illustrate our results on a phasic bursting neuron model, showing that our proposed approach provides an accurate and efficient way to capture neural dynamics.

%%%%%%%%%%%%%%%%%%%%%%%%%%%%%%%%%%%%%%%%%%%%%%%%%%%%%%%%%%%%%%%%%%%%%%%%%%%%

\section*{Introduction}

Connecting single-neuron spiking to the collective dynamics that emerge in neural populations remains a central challenge in systems neuroscience. As well as representing a major barrier in our understanding of fundamental neural function, this challenge has recently acquired new saliency due to the rapid improvements in technologies which can measure neural population activity {\it in vitro} and {\it in vivo} at unprecedented temporal and spatial resolution \citep{jun2017fully,maccione2014following}. Such technologies hold immense promise in elucidating both normal neural functioning and the aetiology of many diseases, yet their high dimensionality and complexity pose formidable statistical challenges.
In response to these needs, recent years have seen considerable efforts to develop strategies for extracting and modeling information from large-scale spiking neural recordings. Two of the most successful strategies that emerged in the last decade are latent state-space models (SSMs), and autoregressive point-process generalized linear models (PPGLMs). 

Latent state-space models describe neural spiking as arising from the unobserved latent dynamics of an auxiliary intensity field, which can model both internal and external factors contributing to the dynamics (e.g. \citealt{macke2015estimating,sussillo2016lfads,zhao2016interpretable,byron2009gaussian,smith2003estimating}). Mathematically, such models generally take the form of a Cox process \citep{kingman1993poisson} where the intensity field obeys some (discrete or continuous time) evolution equations. This representation therefore recasts the analysis of spike trains within a well-established line of research in statistical signal processing, leveraging both classical tools and more recent developments (e.g. \citealt{smith2003estimating,wu2017gaussian, gao2016linear, pfau2013robust, zhao2016variational,surace2017avoid}). These models have been used in a variety of tasks, such as describing population spiking activity in the motor system (e.g. \citealt{aghagolzadeh2014latent,aghagolzadeh2016inference,
churchland2012neural,michaels2017neural}). However, while such models can certainly lead to biological insights, latent state-space models remain \emph{phenomenological}: the recurrent spiking activity itself does not implement the latent state-space dynamics (Fig. \ref{fig:modelcompare}B).

Autoregressive PPGLM models (Fig. \ref{fig:modelcompare}A) treat spiking events from neurons as \emph{point events} 
arising from a latent inhomogeneous Poisson process \citep{truccolo2005point, truccolo2010collective, truccolo2010stochastic, truccolo2017point}. To fit such models, Generalized Linear
Model (GLM) regression is used to map observed spiking events to both extrinsic variables, like stimuli or motor output, and intrinsic spiking history \citep{truccolo2005point}. PPGLM models are especially useful for statistical tests on sources of variability in neural spiking (e.g. \citealt{rule2015contribution,rule2017dissociation}), and benefit from a simple fitting procedure that can often be solved by convex optimization. However, they may require careful regularization to avoid instability \citep{hocker2017multistep, gerhard2017stability}, and can fail to generalize outside of regimes in which they were trained \citep{weber2017capturing}. Importantly, PPGLM models suffer from confounds if there are unobserved sources of neural variability \citep{lawhern2010population}. This is especially apparent when the recorded neural population is a small subsample of the population, and latent state-space models can be more accurate in decoding applications 
\citep{aghagolzadeh2016inference}. 

\begin{figure}
\centering{
\includegraphics[width=0.8\textwidth]{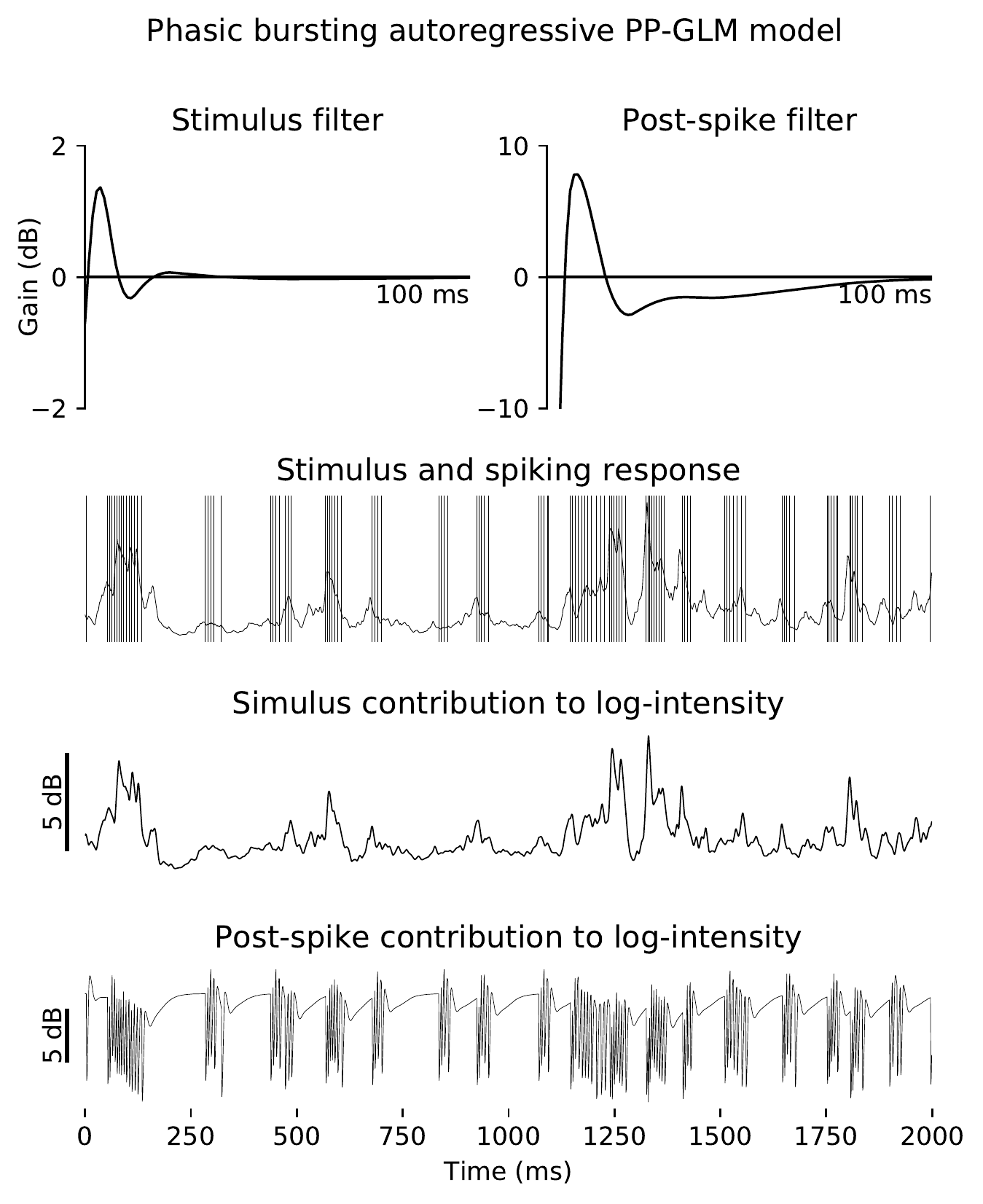}
}
\caption{\emph{Phasic bursting neuron as emulated by an autoregressive PPGLM model.}
Trained on spiking output from Izhikevich neuron model with parameters
$a,b,c,d,dt = 0.02,0.25,-55,0.05,1.0$ \citep{weber2017capturing}. Training data were sampled using Poisson noise stimulation (mean=variance) with current I=0.5 filtered through an alpha-function synapse with time constant $\tau=10$ ms. Testing stimuli (as plotted) were generated by sampling Poisson noise (mean=0.5) from a 5 Hz low-pass log-Gaussian process. 
}
\label{fig:phasicburst}
\end{figure}

In this letter, we establish a mathematical connection between autoregressive PPGLM models and SSMs based on low dimensional, low-order approximation to an exact infinite-dimensional representation of a PPGLM. Unlike previous work, which explored mean-field limits \citep{gerhard2017stability,chevallier2017mean,galves2015modeling,delarue2015particle}, we use Gaussian moment-closure (e.g. \citealt{schnoerr2015comparison,schnoerr2017approximation}) to capture the excitatory effects of fluctuations and process autocorrelations. In doing so, we convert the auto-history effects in spiking into nonlinear dynamics in a low-dimensional latent state space. This converts an autoregressive point-process into a latent-variable Cox process, where spikes are then viewed as Poisson events driven by latent states. This connection, as well as being interesting in its own right, also provides a valuable cross-fertilization opportunity between the two approaches. 
For example, the issue of runaway self-excitation in PPGLMs emerges as divergence in the moment closure ordinary differential equations, leading to practical insights into obtaining a stabilized state-space analogue of the autoregressive PPGLM.
We illustrate the approach on the case study of the phasic bursting of an Izhikevich \citep{izhikevich2003simple} neuron model (Figure \ref{fig:phasicburst}) considered in \citet{weber2017capturing}, showing that our approach achieves both high accuracy in the mean and can capture remarkably well the fluctuations of the process. 

\section*{Results}
We start by recapitulating some basic notations and definitions from both PPGLMs and SSMs. We then provide a detailed derivation of the mathematical connection  between the two frameworks, highlighting all the approximations we make in the process. We finally illustrate the performance of the method in an application case study.

\subsection*{Point-process Generalized Linear Models (PPGLM)} 

A point process (PP) is a subset of dimension zero of a higher-dimensional space \citep{brillinger1988maximum,truccolo2005point}. For our purposes, we will only consider PPs over the time domain, so we will equivalently consider a realization of a PP as a series of points in time $y(t)$, where each point (spiking event) is a delta distribution at the event time. We can associate with a PP in time a locally constant counting process $N(t)$  that counts the cumulative number of events up to time $t$. The process $y(t)$ can be thought of as representing the spike train output by a neuron, while the cumulative process $N(t)$ provides a clearer notation for the derivations that follow:
\begin{equation*}
\begin{aligned}
N(t)   &= \text{ \# events $\le$ t }  
\\y(t) &= \tfrac {d}{dt} N(t) = \sum_{\tau\in\text{events}} \delta(t{=}\tau).
\end{aligned}
\end{equation*}
We restrict our attention to PPs that can be described by an underlying intensity function $\lambda(t)$; in the simplest case, event counts between times $t_1$ and $t_2$ occur with a Poisson distribution, with a mean rate given by the integral of $\lambda(t)$ over the time window.
\begin{equation}
\Pr\left(\textstyle N(t{+}\Delta)-N(t) = k\right) \sim \operatorname{Poisson}\left(\textstyle\int_{t}^{t+\Delta} \lambda(t) dt\right)
\end{equation}
In the autoregressive PPGLM (Fig. \ref{fig:modelcompare}A), one models the  intensity $\lambda(t)$ conditioned on the past events, as well as extrinsic covariates $x(t)$. Generally,
\begin{equation}
f\left(\lambda(t)\right) = \mu + F^\top x(t) + \textstyle\int_0^{\infty} H(\tau) y(t-\tau) d\tau
\label{eq:glm}
\end{equation}
where $f$ is called the link function, $F$ is a matrix or operator projecting extrinsic covariates down to the dimensionality of the point-process, $\mu$ is a mean or bias parameter, and $H$ is a history filter function. The inputs and bias $\mu + F^\top x(t)$ are fixed, and can be denoted by a single time-dependent input function $I(t) = F^\top x(t) + \mu$. 
Here we will take the link function $f$ to be the natural logarithm. Re-writing Eq. \ref{eq:glm}, making time-dependence implicit where unambiguous, and denoting $\textstyle\int_0^{\infty} H(\tau) y(t-\tau) d\tau$ as $H^\top y$, we will explore generalized linear models of the form:
\begin{equation}
\lambda = \exp\left(\mu + F^\top x + H^\top y\right)
\end{equation}
Autoregressive PPGLMs can emulate various firing behaviors of real neurons \citep{weber2017capturing}. For example phasic bursting neurons (Fig. \ref{fig:phasicburst}) exhibit complex autohistory dependence on both fast and slow timescales. This dependence of the process on intrinsic history confers additional slow dynamics.

\subsection*{Latent state-space point-process models}

An alternative strategy for capturing slow dynamics in neural spike trains is to postulate a slow, \emph{latent} dynamical system responsible for bursting and post-burst inhibition. This approach is taken by latent state-space models (SSMs), which are often viewed as functionally distinct from autoregressive PPGLMs (Fig. \ref{fig:modelcompare}B). 

In general, a latent state-space model (SSM) describes how both deterministic and stochastic dynamics of a latent variable $x$ affect the intensity $\lambda$ of a point process:
\begin{equation}
\begin{aligned}
dx(t) &= u(x,t) dt + \sigma(x,t) dW
\\\lambda &= v(x,t)
\\dN &\sim \text{Poisson}(\lambda\cdot dt),
\end{aligned}
\end{equation}
where $dW$ is the derivative of the standard Wiener process, reflecting fluctuations. The functions $u$, $\sigma$, and $v$ describe, respectively, deterministic evolution, stochastic fluctuations, and the observation model. In the case of, for example, the Poisson Linear Dynamical System (PLDS; \citealt{macke2011empirical}), the latent dynamics are linear with fixed Gaussian noise:
\begin{equation}
\begin{aligned}
dx(t) &= A x + w(t) + \sigma dW
\\\lambda &= \exp\left(\mu + F^\top x + H^\top y\right)
\\dN &\sim \text{Poisson}(\lambda\cdot dt),
\end{aligned}
\end{equation}
where $w(t)$ reflects inputs into the latent state-space. Latent state-space models of point-processes have been investigated in detail (e.g. \citealt{macke2011empirical,smith2003estimating}), and mature inference approaches are available to estimate states and parameters from data \citep{lawhern2010population,macke2015estimating,buesing2012spectral,rue2009approximate,cseke2016sparse}. However, such models are typically phenomenological, lacking a clear physiological interpretation of the latent dynamics. Importantly, the point-process history is typically fit as if it were another extrinsic covariate, and the effects of Poisson fluctuations are either neglected or handled in mean-field limit. This obscures the dynamical role of population spiking history and its fluctuations in the autoregressive PPGLM. In the remainder of this paper, we illustrate that the history dependence of PPGLM models implicitly defines a latent-state space model over moments of the process history.

\subsection*{The auxiliary history process of a PPGLM}
 
The (possibly infinite) history dependence makes autoregressive PPGLMs non-Markovian dynamical systems (c.f. \citealt{truccolo2017point} Eq. 6). However, a crucial insight is that, since the dependence on history is linear, we can re-interpret the history dependence as a linear filter in time, and approximate its effect on the conditional intensity using a low-dimensional linear dynamical system.

To formalize this, let us introduce an auxiliary history process $h(\tau,t)$ that ``stores" the history of the process $y(t)$. One can view $h(\tau,t)$ as a delay-line that tracks the signal $y(t)$. The time evolution of $h$ is given by:
\begin{equation}
\partial_t h(\tau,t) = -\partial_\tau h(\tau,t) + \delta_{\tau=0} dN(t)
\label{eq:delayline}
\end{equation}
where $\delta_{\tau=0}$ indicates that new events $y(t)=dN(t)$ should be inserted into the history process at $\tau{=}0$, and $\partial_\tau$ is the derivative with respect to time lag $\tau$. This converts the autoregressive PPGLM to a stationary Markovian process over an augmented (infinite dimensional) state space:
\begin{equation}
\begin{aligned}
dN(t)&\sim\operatorname{Poisson}(\lambda \cdot dt)
\\\lambda(t) &= \exp\left( H(\tau)^\top h(\tau,t) + I(t)\right)
\\\partial_t h(\tau,t) &= \delta_{\tau=0} dN(t) - \partial_\tau h(\tau,t)
\end{aligned}
\label{eq:spde}
\end{equation}
where $H(\tau)$ is the history filter introduced in equation \eqref{eq:glm}. In this formulation, the history process $h(\tau,t)$ is still a point-process. However, the interaction between history $h(\tau,t)$ and the intensity $\lambda(t)$ is mediated entirely by the \emph{projection} $H(\tau)^\top h(\tau,t)$, which averages over the process history. To capture the relevant influences of the process history, it then suffices to capture the effects of Poisson variability on this averaged projection.

\subsection*{A continuous approximation}

In the limit where events are frequent, the Poisson process $dN(t)$ can be approximated as a Wiener process with mean and variance equal to the instantaneous point-process intensity $\lambda(t)$. In the derivations that follow, we omit explicit notation of time-dependence (e.g. $\lambda(t)$, $h(\tau,t)$) where unambiguous:
\begin{equation}
\begin{aligned}
dN & \approx \lambda dt + \sqrt{\lambda} dW,
\end{aligned}
\end{equation}
This approximation holds when averaging over a population of weakly-coupled neurons, or averaging over slow-timescales of a single neuron. Applying this approximation to the driving noise term in the evolution equation for the auxiliary history process \eqref{eq:spde}, we obtain a continuous (in time, and in state) infinite-dimensional approximation of the PPGLM:
\begin{equation}
\begin{aligned}
d h &= \left(\delta_{\tau=0}\lambda - \partial_\tau h\right)dt + \delta_{\tau=0}\sqrt{\lambda} dW
\\\lambda &= \exp\left( H^\top h + I(t)\right).
\end{aligned}
\end{equation}
Because the dimensionality of the history process $h(\tau,t)$ is infinite, this is a stochastic partial differential equation (SPDE). Importantly, this is a system of equations for the \emph{history} of the process, not the instantaneous rate $\lambda(t)$. This SPDE is analytically intractable due to the exponential link function, however it is possible to derive  an (infinite) set of coupled moment equations for the process; we can then close these equations by setting all cumulants of order greater than two to zero, effectively enforcing Gaussianity of the the history process $h(\tau,t)$. The (exact) equation for the process mean $\mu(\tau) = \left<h(\tau)\right>$ is as follows
\begin{equation}
\begin{aligned}
\partial_t \mu &= \partial_t \left<h\right>
\\&= \left< \delta_{\tau=0}\lambda - \partial_\tau h \right>
\\&= \delta_{\tau=0}\left< \lambda\right> - \partial_\tau \mu.
\end{aligned}
\end{equation}
If we consider a log-linear model, and since $h(\tau, t)$ is approximated as Gaussian, $\lambda$ is log-normally distributed with mean: 
\begin{equation}
\left< \lambda\right>=\exp\left(H^\top \mu + I(t) +\tfrac 1 2 H^\top \Sigma H \right)
\end{equation}
This expectation incorporates second-order effects $\tfrac 1 2 H^\top \Sigma H$ introduced by fluctuations and time correlations mediated through the history filter. Note that the time-evolution of the first moment depends on the covariance $\Sigma(\tau,\tau')$.
The time derivative of the covariance has both deterministic and stochastic contributions. Overall, the deterministic contribution to the derivative of the covariance can be written as $J \Sigma + \Sigma J^\top$, where (see Appendix A) 
\begin{equation}
J = \delta_{\tau=0} \left<\lambda\right> H^\top -\partial_{\tau}.
\end{equation}
The covariance also has a noise contribution from the $dW$ term, with variance proportional to the expected firing rate $Q = \delta_{\tau=0} \left<\lambda\right> \delta_{\tau=0}^\top$. In sum, the derivative of the second moment is:
\begin{equation}
\begin{aligned}
\partial_t \Sigma &= J\Sigma + \Sigma J^\top + Q
\end{aligned}
\end{equation}
This notation resembles continuous-time Kalman-Bucy filter \citep{kalman1961new}, for which $J(t)$ would be a Jacobian of the mean update, and $Q(t)$ would reflect the system noise.

%%%%%%%%%%%%%%%%%%%%%%%%%%%%%%%%%%%%%%%%%%%%%%%%%%%%%%%%%%%%%%%%%%%
\subsection*{Finite dimensional projection}

\begin{figure}[H]
\centering{
\includegraphics[width=\textwidth]{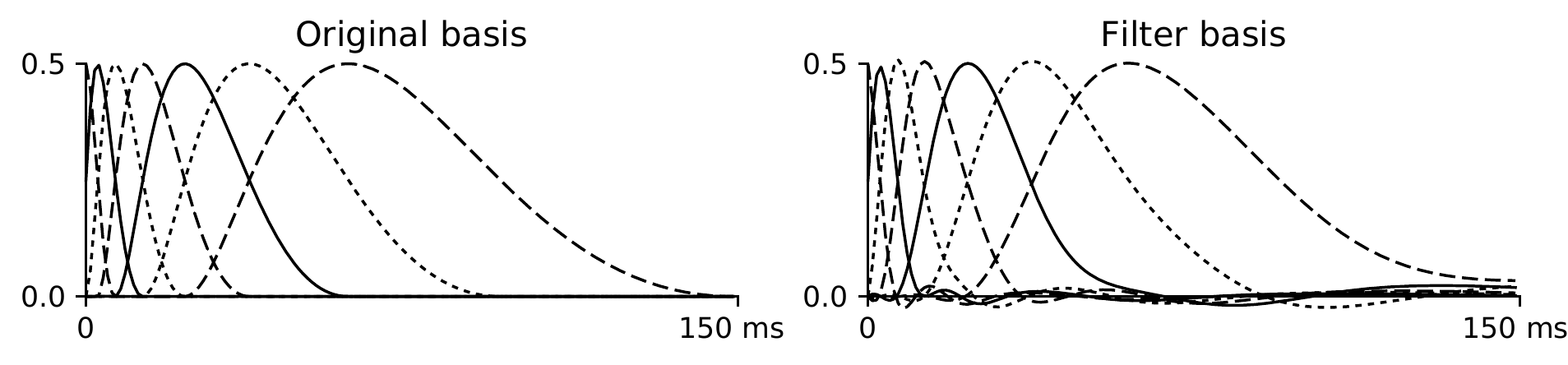}}
\caption{\emph{Basis projection of the history process yields a linear filter approximating the original history basis elements.}
\emph{Left:}
History dependence in autoregressive point-process models is typically
regularized by using a finite history basis.
\emph{Right:}
To convert history basis elements into a linear dynamical system, one projects the infinite-dimensional delay-line (Eq. \ref{eq:delayline}) onto the low-dimensional basis. The resulting linear system has response functions that approximate the history basis. Note, however, the ringing introduced by the approximation.}
\label{fig:basis}
\end{figure}

The history process $h(\tau,t)$ is infinite dimensional. To make inference and simulation practical,  one represents the continuous history filter $H(\tau)$ by a finite collection of basis functions $B(\tau) = \left\{B_1(\tau),..,B_K(\tau)\right\}$ (Fig. \ref{fig:basis}). A common choice is to use a cosine basis, for example from \citet{weber2017capturing}:
\begin{equation*}
B_j(t) = \tfrac 1 2 \cos(a \log[t+c] - \phi_j) + \tfrac 1 2
\end{equation*}
Where parameters $a$ and $c$ select the base and offset of the functions in log-time, respectively, and $\phi_j$ are offsets in integer multiples of $\pi/2$. This basis projection moves us from an infinite dimensional history $h(\tau,t)$ to a finite state space $z(t)=\{z_1(t),..,z_k(t)\}$ defined by the projection $B(\tau) = \left\{B_0(\tau),..,B_k(\tau)\right\}$ of $h(\tau,t)$. 
\begin{equation}
z_i(t)= \textstyle\int B_i(\tau) h(\tau,t) d\tau
\label{eq:basis1}
\end{equation}
$B$ should be normalized so that volume is preserved at every time $\tau$, i.e. $\forall\tau$, $\textstyle\sum_i B_i(\tau){=}1$, so that the history basis features can be treated as Poisson random variables. In practice the history will not extend for infinite time, and the final basis functions may be omitted. The continuous history filter $H(\tau)$ is replaced by discrete weights $\beta_i{=}\int_\tau H(\tau) B_i(\tau)$:
\begin{equation}
\begin{aligned}
H(\tau)^\top h(\tau,t) &= \textstyle\int_0^\infty H(\tau) h(\tau,t) d\tau 
\\&\approx\textstyle\sum_i \beta_i \textstyle\int B_i(\tau) h(\tau,t) d\tau.
\end{aligned}
\label{eq:basis2}
\end{equation}
The time-evolution of $z(t)$ can be written in terms of $h(\tau,t)$:
\begin{equation}
\begin{aligned}
\partial_t z(t) &= \partial_t B h(\tau,t) 
\\&= B \partial_t h(\tau,t) 
\\&= -B\partial_\tau h(\tau,t) + B \delta_{\tau=0} y(t)
\end{aligned}
\label{eq:basis3}
\end{equation}
We can approximately recover the state of the delay line $h(\tau,t)$ from the basis projection 
using the Moore-Penrose pseudoinverse of the basis $B^{+}$:
\begin{equation}
h(\tau,t) \approx B^{+} z(t) = \textstyle\sum_i z_i(t) B_i(t-\tau)
\label{eq:basis4} 
\end{equation}
This yields a closed approximate dynamical system for computing the convolution of the history basis $B$ with a signal $y(t)$
\begin{equation}
\partial_t z \approx \partial_t \tilde z = -B\partial_\tau B^{+} \tilde z + B \delta_{\tau=0} y(t)
\label{eq:basis5}
\end{equation} 
This is a finite-dimensional linear system $\tilde z$ that approximates the history using basis projection.
\begin{equation}
\begin{aligned}
\partial_t \tilde z &= C y(t) - A \tilde z ,\qquad
\\A &= B\partial_\tau B^{+}  
\\C &= B \delta_{\tau=0}
\end{aligned}
\label{eq:basis6}
\end{equation}
\emph{In silico}, the differentiation $\partial_\tau$ and Dirac delta $\delta_{\tau=0}$ operators are implemented as matrices representing the discrete derivative and a point mass over one time-step, respectively. The above basis projection then yields low-dimensional linear operators defining a dynamical system. The resulting process is:
\begin{equation}
\begin{aligned}
y(t)&\sim\operatorname{Poisson}(\lambda)
\\\lambda(t) &= \exp\left( \beta^\top  \tilde z(t) + I(t) \right)
\\\partial_t \tilde z(t) &= C y(t) - A  \tilde z(t)
\end{aligned}
\label{eq:basis7}
\end{equation}

The basis projections integrate over an extended time-window. If intensity $\lambda(t)$ is approximately constant during this time window, then the basis-projected history variables $z(t)=(z_1(t),..,z_k(t))$ are Poisson variables with rate and variance $z_i(t)$. These projections, by virtue of integrating over longer timescales, can be approximated as Gaussian. Fluctuations that are far from Gaussian in the point process $y(t)$ can be well approximated as Gaussian (with mean equal to variance) projections of the history process. In this case, we may approximate the Poisson process as a Wiener process that is continuous in time:
\begin{equation}
\begin{aligned}
d z(t) &= \left[C \lambda(t) - A z(t)\right] dt + C\sqrt{\lambda(t)} dW
\\\lambda(t) &= \exp\left( \beta^\top z(t) + I(t) \right)
\end{aligned}
\label{eq:basis8}
\end{equation}

\begin{figure}[p!]
\centering{ 
\includegraphics[width=\textwidth]{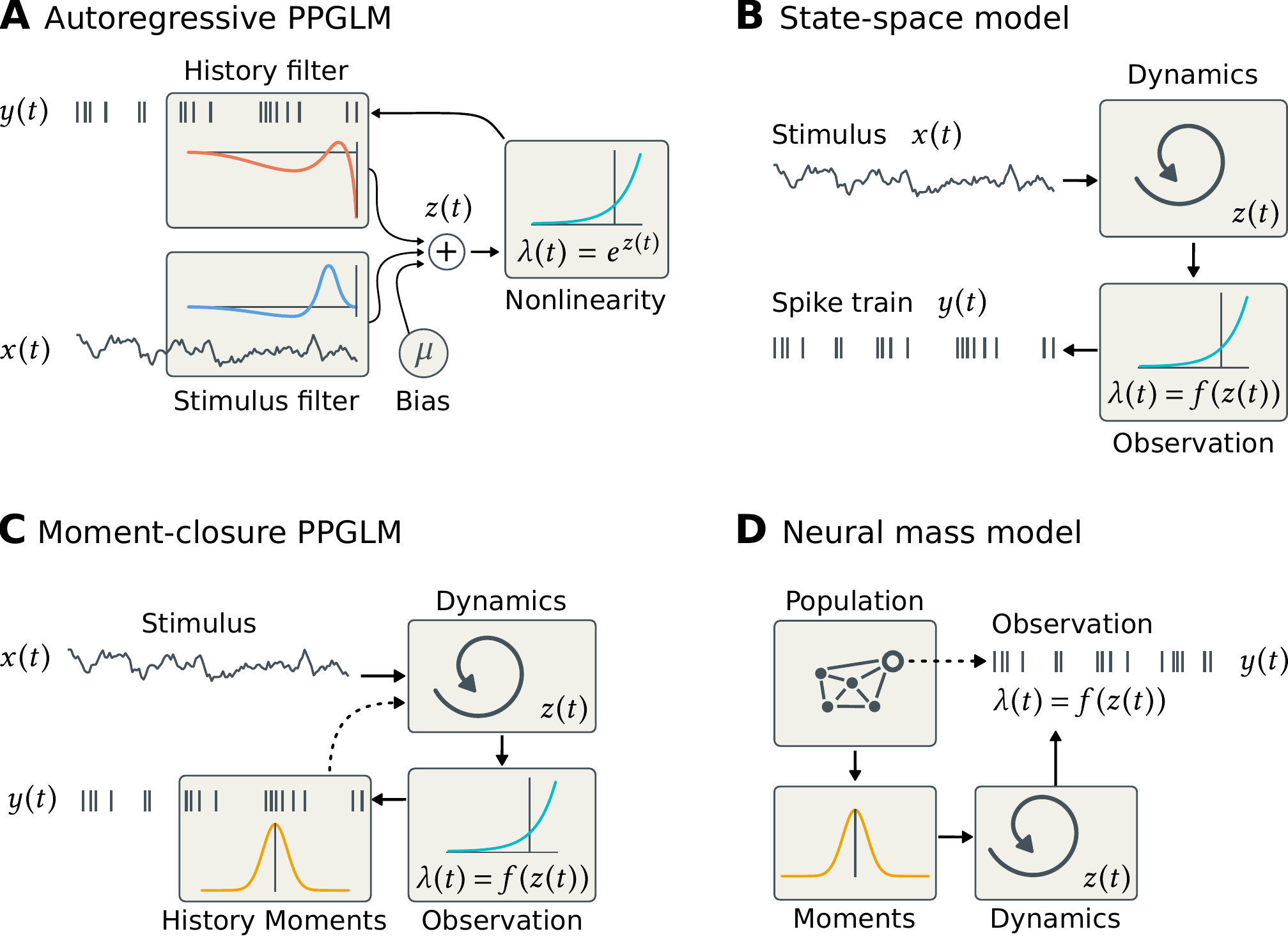}}  
\caption{\emph{Moment-closure of autoregressive PPGLMs combines aspects of three modeling approaches.}
\textbf{A} Log-linear autoregressive PPGLM framework (e.g. \citealt{weber2017capturing}). Dependence on the history of both extrinsic covariates $x(t)$ and the process itself $y(t)$ are mediated by linear filters, which are combined to predict the instantaneous log-intensity of the process. 
\textbf{B}
Latent state-space models learn a hidden dynamical system, which can be driven by both extrinsic covariates and spiking outputs. Such models are often fit using expectation-maximization, and the learned dynamics are descriptive.
\textbf{C}
Moment-closure recasts autoregressive PPGLM models as state-space models, where the latent dynamical state space has a physical interpretation as moments of the process history.
\textbf{D}
Compare to neural mass and neural field models, which define dynamics on a state-space with a physical interpretation as moments of neural population activity. 
}
\label{fig:modelcompare}
\end{figure}

Analogously to the moment-closure for the infinite-dimensional system, one can derive a Gaussian moment-closure for the low-dimensional basis-projected system. The equations for the evolution of the mean and second moment in the finite basis projection are:
\begin{equation}
\begin{aligned} 
\\ \partial_t \mu_z &=  - A\mu_z + C \left< \lambda \right>
\\ {\left<\lambda\right>} &= \exp\left( \beta^\top  \mu_z + I(t) + \tfrac 1 2 \beta^\top \Sigma_z \beta \right)
\\ \partial_t \Sigma_z &= J \Sigma_z + \Sigma_z J^\top + Q(t)
\\ J &= C \left< \lambda \right> \beta^\top - A
\\ Q &= C \left< \lambda \right> C^\top
\end{aligned}
\label{eq:basis9}
\end{equation}
Equations \eqref{eq:basis9} {are reminiscent of} classical neural mass and neural field models (\citealt{amari1975,amari1977,amari1983,wilson1972}; e.g. Fig. \ref{fig:modelcompare}D). Unlike neural field models, however, the moment equations \eqref{eq:basis9} do not arise from population averages, but rather by directly considering the expected behavior of the stochastic process describing the neural spike train (Fig. \ref{fig:modelcompare}C).

It is worth reflecting more on this analogy, and on the limitations of this representation.
Spiking events are a dramatic all-or-nothing events that cannot be approximated by a continuous stochastic process . Accordingly, one would expect the finite-dimensional moment closure system to fail to capture rapid fluctuations. However, for slow timescales, this Gaussian approximation can be accurate \emph{even for a single neuron}. In contrast to the neural field interpretation, which averages over a large population at each time instant, one can average over an extended time window, and arrive at an approximation for slow timescales (e.g. Fig. \ref{fig:burst}). 
A pictorial description of the relationship of the proposed moment closure approach to PPGLMs, SSMs and neural field models is summarized in Figure~\ref{fig:modelcompare}.

\subsection*{Case study: the Izhikevich neuron model}

\begin{figure}[H]
\centering{ 
\includegraphics[width=\textwidth]{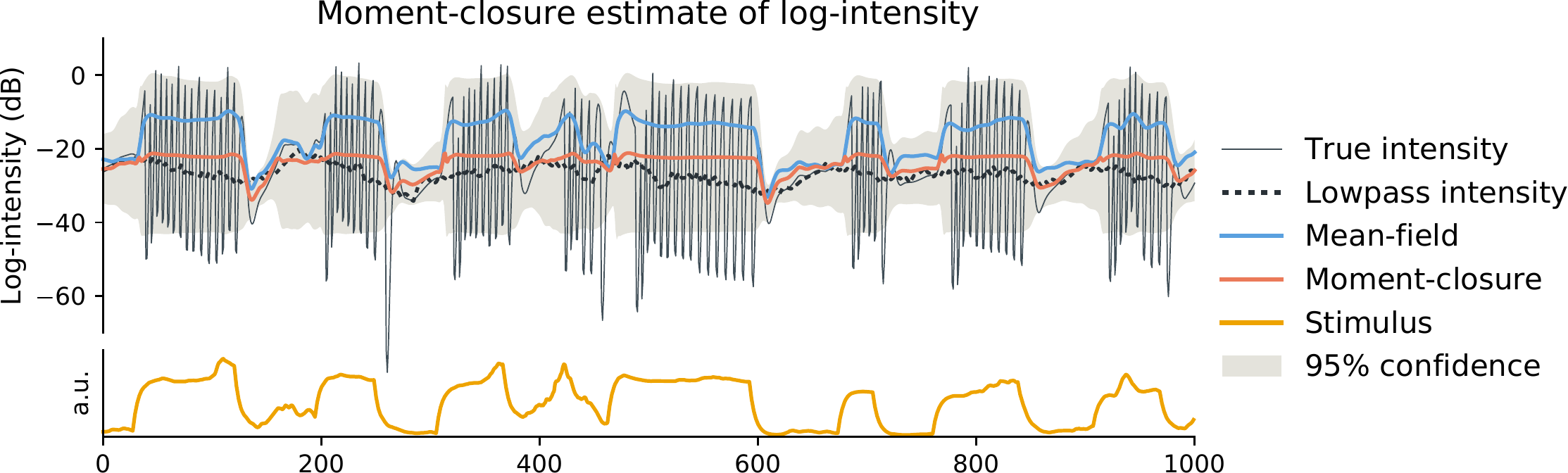}}  
\caption{\emph{Moment closure captures slow timescales in the mean, and fast timescales in the variance.}
Moment-closure (red line) tracks the low-frequency (dashed line; 50~ms boxcar smoothed) trends in the spiking log-rate (thin line). Bursts of spiking are captured as an increase in variance (95\% = 1.96$\sigma$, shaded). Compare to mean-field (blue line), which cannot account for fluctuation effects and time correlations, and over-estimates the mean log-intensity during bursts.  
Model was trained to a phasic bursting neuron \citep{weber2017capturing}, 
using shuffled square pulse stimuli between 10 and 500~ms and between 0.1 and 3.0~mV in amplitude. To improve fit robustness, log-Ornstein-Uhlenbeck (OU) noise was added to the training stimulus with (log) mean of -1, steady-state variance of 0.75, and time constant of 50 ms.
}
\label{fig:burst}
\end{figure}

We consider the effectiveness of this approach on the case study of a PPGLM emulation of the Izhikevich neuron model, considered in \citet{weber2017capturing}. We compare the accuracy of the Gaussian moment closure with a mean field approach (Appendix B). 
Figure \ref{fig:burst} illustrates moment-closure of a phasic-bursting Izhikevich neuron emulated with a PPGLM (Fig. \ref{fig:phasicburst}). By averaging over the history process, slow-timescales in the autoregressive point-process are captured in the Gaussian moment-closure. Unlike a mean-field model, which considers the large-population limit of weakly-coupled neurons, moment-closure is able to capture the influence of Poisson variability on the dynamics. 

Additionally, mean-field considers only a single path in the process history, whereas Gaussian moment-closure provides an approximation for a \emph{distribution} over paths, with fluctuations and autocorrelations taken into account. This has the benefit that the moment-closure system is sensitive to the combined effects of self-excitation and Poisson fluctuations, and captures, for example, the self-excitation during a burst using the second-order second moment terms. This reveals another benefit of the moment-closure approach: runaway self-excitation \citep{hocker2017multistep, gerhard2017stability, weber2017capturing} is detected in the moment-closure as a divergence of the mean or second moment terms. This self-excitation, however, introduces some numerical challenges.

Typically, the post-spike filters in PPGLM models confer large, rapid negative-feedback at fast timescales, in order to model the absolute refractoriness of spike trains. The combination of large negative feedback at fast timescales, and emergent dynamics at slow timescales, make the state-space moment equations of the GLM stiff: small time-steps must be taken to capture the refractory effects, even when only slow-timescales are relevant to the model. We note that a second-order approximation to the moment equations is less stiff and exhibits reduced runaway self-excitation (Appendix B). This highlights a major benefit of the moment closure approach: numerical issues which prove difficult or intractable in the original GLM representation can be more readily addressed in a state-space model.
Importantly, the basis-projects moment-closure system is an ordinary differential equation with a form reminiscent of nonlinear (extended) continuous-time Kalman-Bucy filtering, and the moment-closure state-space equations allow tools for reasoning about the stability of ordinary differential equations to be applied to PPGLMs.

\section*{Discussion}

In this letter, we have introduced a mathematical connection between PPGLMs and SSMs that provides an explicit, constructive procedure to fit neural spike train data.
Autoregressive point-processes and state-space models have been combined before (e.g. \citealt{zhao2016interpretable,smith2003estimating,lawhern2010population,eden2004dynamic}), {but so far always in a manner that treats the latent state-space as an extrinsic driver of neural activity. Importantly, the generative, dynamical effects of Poisson fluctuations and process autocorrelations are not directly addressed in previous approaches.}  Additionally, although PPGLM models can condition on population spiking history during training, this conditioning addresses only a single sample-path of the process history, and does not reflect a recurrent dynamical model in which spiking outputs and their fluctuations lead to the emergence of collective dynamics. The moment-closure approach outlined here can be used to incorporate auto-history effects into a latent state-space model, where conditioning on the process history is replaced by a Bayesian filtering update which updates the moments of the history process at each time-step.

Our results highlight the capacity of PPGLM models to implicitly learn hidden causes of neural firing through the autoregressive history filter. For example, a collective network mode at 20 Hz may induce spiking rhythmicity that is detected in the point-process history filter, even if isolated neurons do not exhibit this oscillation. The history dependence of autoregressive PPGLMs defines a latent variable process that captures both process auto-history effects and the influence of unobserved hidden modes or inputs. The moments of the population spiking history can be identified with the latent variables explaining population spiking. This interpretation replaces pairwise coupling in a neural population in the PPGLM formulation with a coupling of single neurons to a shared latent-variable history process.

The identification of latent states with moments of the population history opens up a new interpretation connecting both PPGLMs and latent state-space models to neural field models, a rich area of research in theoretical neuroscience \citep{amari1975,amari1977,amari1983,wilson1972}. It suggests that under some conditions, neural-field models may be interpreted as latent variable models and fit using modern techniques for latent state-space models. Conversely, this new connection illustrates that some latent-state space models may be viewed not only as modeling hidden causes of spiking, but also as capturing statistical moments of the population that are relevant for neural dynamics in a neural-field sense. The precise convergence of a moment closure PPGLM to a neural field model remains to be better explored mathematically. Convergence of moment closure approximations has been studied extensively in the area of stochastic chemical reactions \citep{schnoerr2014validity,schnoerr2015comparison,schnoerr2017approximation}. Indeed, our approach was partly inspired by recent work on chemical reaction-diffusion systems \citep{schnoerr2016}, in which pairwise interactions between pointwise agents in space are replaced by a coupling of single agents to a statistical field. In contrast to chemical reaction systems however, we model self-interactions of a point-process over time, and capture also the effects of fluctuations on the system.

There are two major benefits of the moment closure representation of PPGLM models. First, autoregressive time-dependencies are converted to a low-dimensional system of ordinary differential equations, re-interpreting the PPGLM as a dynamical latent-state space model of a similar form as phenomenological latent-dynamics models and population neural field models. Second, moment closure equations open up new strategies for estimating PPGLM models. A major challenge to fitting PPGLM models to large populations is the challenges in estimating a large number of pairwise interactions. Our work suggests a different avenue toward estimating such large models: a low-dimensional latent-variable stochastic process with a suitable nonlinearity and Poisson noise can be interpreted as a process governing the the moments of an PPGLM model. This allows the extensive methodological advancements toward identifying low-dimensional state-space models to be applied to autoregressive point-processes.
This could be especially useful in systems in which the spiking output, and fluctuations therein, substantially influences the population dynamics.

Another challenge in estimating PPGLM models is ensuring that the fitted model accurately captures dynamics \citep{hocker2017multistep, gerhard2017stability}. The moment-closure equations outlined here allow process moments to be estimated, along with model likelihood, using Bayesian filtering. In addition to filtering over a distribution of paths in the process history, filtering can also average over models, and thus implicitly capture both fluctuation effects and model uncertainty. However, it remains the subject of future work to apply the moment-closure approach in inference. Other methods, such as particle filtering, may be useful in situations where the latent state-space distribution is highly non-Gaussian. 

\subsection*{Acknowledgments}
The authors acknowledge support from the Engineering and Physical Sciences Research Council under grant EP/L027208/1 \emph{Large scale spatio-temporal point processes: novel machine learning methodologies and application to neural multi-electrode arrays}
\section*{Appendix}

\subsection*{Appendix A: time evolution of the second moment}
We work with the time evolution of the covariance $\Sigma$ rather than the second moment $\left<h h^\top \right>$, for improved stability in numerical implementations. 
\begin{equation}
\Sigma = \left<h h^\top \right> - \left<h\right>\left<h\right>^\top
\end{equation}
Differentiating the covariance:
\begin{equation}
\begin{aligned}
\partial_t \Sigma &= \partial_t \left( \left<h h^\top \right> - \left<h\right>\left<h\right>^\top \right)
\\
&= \partial_t \left<h h^\top \right> - \partial_t \left(\left<h\right>\left<h\right>^\top \right)
\\
&= \left<(\partial_t h) h^\top \right>
+\left<h (\partial_t h^\top) \right> 
-\left(\partial_t \left<h\right>^{\vphantom{\top}} \right) \left<h\right>^\top
-\left<h\right> \left(\partial_t\left<h\right>^\top \right)
\end{aligned}
\end{equation}
This expression consists of two sets of symmetric terms arising from the product rule. Examine one set of terms, and substitute in the delay-line evolution Eq. \ref{eq:delayline}:
\begin{equation}
\begin{aligned}
\left<(\partial_t h) h^\top \right> - \left(\partial_t \left<h\right>^{\vphantom{\top}} \right) \left<h\right>^\top
&=
\left<[\delta_{\tau=0} \lambda - \partial_\tau h] h^\top \right>-\left[\delta_{\tau=0} \left<\lambda\right> - \partial_\tau \left<h\right>\right] \left<h\right>^\top
\\&=
\delta_{\tau=0} \left[\left< \lambda h^\top \right>- \left<\lambda\right>\left<h\right>^\top\right]-\partial_\tau \left[\left< h h^\top \right>-\left<h\right>\left<h\right>^\top \right]
\end{aligned}
\label{eq:dcov}
\end{equation}
This expression is linear in the first two moments, except for the expectation $\left< \lambda h^\top \right>$. This expectation is taken over the Gaussian history process with mean $\left<h\right>$ and covariance $\Sigma$, and can be computed by completing the square using $m{=}\left<h\right>{+}\Sigma H$ in the Gaussian integral:
\begin{equation}
\begin{aligned}
\left< \lambda h^\top \right> 
&=\left< h^\top e^{H^\top h + I } \right>
\\ &=e^{I(t)} \textstyle \int_{dh} h e^{H^\top h}
\frac 1 {\sqrt{|2\pi\Sigma|}} e^{- \frac 1 2 (h-\left<h\right>)^\top \Sigma^{-1} (h-\left<h\right>)}
\\ &= e^{I(t)} e^{\tfrac 1 2 (m^\top \Sigma^{-1} m - \left<h\right>^\top \Sigma^{-1} \left<h\right>)}\cdot m^\top
\\ &= e^{H^\top \left<h\right> + I(t) + \tfrac 1 2 H^\top \Sigma H} \cdot m^\top
\\ &= \left<\lambda\right> \left(\left<h\right> + \Sigma H\right)^\top.
\end{aligned}
\end{equation}
Substituting the above expression into Eq. \ref{eq:dcov} and simplifying yields the deterministic contribution to the evolution of the covariance:
\begin{equation}
\begin{aligned}
\left<(\partial_t h) h^\top \right> - \left(\partial_t \left<h\right>^{\vphantom{\top}} \right) \left<h\right>^\top
&=
\delta_{\tau=0} \left(\left< \lambda \right> \left(\left<h\right> + \Sigma H \right)^\top- \left<\lambda\right>\left<h\right>^\top\right)-\partial_\tau \Sigma
\\&=
\underbrace{
\left(\delta_{\tau=0} \left< \lambda \right> H^\top -\partial_\tau \right)}_{J}\Sigma
\end{aligned}
\end{equation}

\subsection*{Appendix B: the linear noise approximation and a stabilized moment closure}

A simpler alternative to moment-closure is the Linear Noise Approximation (LNA), which uses a deterministic mean $\bar\lambda$ obtained in the limit of a large, weakly-coupled population for which the effect of fluctuations on the mean is negligible. In the basis-projected system (Eqs. \ref{eq:basis1}-\ref{eq:basis8}) the deterministic mean is:
\begin{equation}
\begin{aligned}
\partial_t \mu_z &= C \bar \lambda - A \mu_z
\\ \bar\lambda &= \exp\left(\beta^\top \mu_z + I(t) \right)
\end{aligned}
\label{eq:mf}
\end{equation}
The LNA describes the second moment as a function of this mean, but does not correct for the influence of fluctuations on the evolution of the mean. The LNA about the deterministic $\bar\lambda$ is:
\begin{equation}
\begin{aligned}
\\ \partial_t \Sigma_z &= J\Sigma + \Sigma J ^\top + C \bar \lambda C^\top
\\ J &= C \bar\lambda \beta^\top + A
\end{aligned}
\label{eq:lna}
\end{equation}
In contrast, moment-closure equations capture an approximate distribution over paths in the history of the process, and can correct for effects of fluctuations on the mean-rate. However, in practice, the moment equations can be stiff and challenging to integrate. This is because self-excitation is typically stabilized by rapid negative-feedback on short timescales, and the exponential nonlinearity can lead to runaway self-excitation. A modified version, which approximates the variance correction to second order, incorporates variance corrections with improved stability:  
\begin{equation}
\begin{aligned}
   \partial_t \mu_z &= C \tilde \lambda - A \mu_z
\\ \partial_t \Sigma_z &= J\Sigma + \Sigma J ^\top + C \tilde \lambda C^\top
\\ J &= C \bar\lambda \beta^\top + A
\\ \bar\lambda &= \exp\left(\beta^\top \mu_z + I(t) \right)
\\ \tilde\lambda &= \bar\lambda \cdot (1 + \tfrac 1 2 \beta^\top \Sigma \beta )
\end{aligned}
\label{eq:secondorder}
\end{equation}
In this system, the mean-field $\bar \lambda$ is used for the deterministic evolution of the variance, and fluctuation effects on the mean rate are approximated at second-order as $\tilde\lambda$. These equations are a heuristic, but in practice provide a state-space analogue of an autoregressive PPGLM that is more stable.

\bibliography{\string./ARPPGLMCox_I_arXiv}{}

\end{document}